\documentclass{Interspeech}

\interspeechcameraready 
\usepackage{comment}
\usepackage{float}
\usepackage{subcaption}
\usepackage{stfloats}
\usepackage{threeparttable}
\usepackage{graphicx}
\usepackage{cite}
\usepackage{multirow}
\usepackage{adjustbox}
\usepackage{pifont}  
\usepackage{multirow, multicol}
\usepackage{makecell}
\usepackage{bbding}
\usepackage{enumitem}
\PassOptionsToPackage{hyphens}{url}
\usepackage{url}
\usepackage{hyperref}

\title{Vision-Integrated High-Quality Neural Speech Coding}

\author{Yao}{Guo}
\author{Yang}{Ai$^*$}
\author{Rui-Chen}{Zheng}
\author{Hui-Peng}{Du}
\author{Xiao-Hang}{Jiang}
\author{Zhen-Hua}{Ling}
\affiliation[nocounter]{National Engineering Research Center of Speech and Language Information Processing}{University of Science and Technology of China}{Hefei, P. R. China}
\email{guoyao1917@mail.ustc.edu.cn, yangai@ustc.edu.cn, zhengruichen@mail.ustc.edu.cn, redmist@mail.ustc.edu.cn, jiang\_xiaohang@mail.ustc.edu.cn, zhling@ustc.edu.cn}

\begin{document}
\maketitle
\renewcommand{\thefootnote}{\fnsymbol{footnote}}
\footnotetext[1]{Corresponding author. This work was funded by Anhui Province Major Science and Technology Research Project under Grant S2023Z20004, the National Nature Science Foundation of China under Grant 62301521 and the Anhui Provincial Natural Science Foundation under Grant 2308085QF200.}
\renewcommand{\thefootnote}{\arabic{footnote}}
\vspace{-1mm}
\begin{abstract}
\vspace{-1mm}    
This paper proposes a novel vision-integrated neural speech codec (VNSC), which aims to enhance speech coding quality by leveraging visual modality information. 
In VNSC, the image analysis-synthesis module extracts visual features from lip images, while the feature fusion module facilitates interaction between the image analysis-synthesis module and the speech coding module, transmitting visual information to assist the speech coding process.
Depending on whether visual information is available during the inference stage, the feature fusion module integrates visual features into the speech coding module using either explicit integration or implicit distillation strategies. 
Experimental results confirm that integrating visual information effectively improves the quality of the decoded speech and enhances the noise robustness of the neural speech codec, without increasing the bitrate.
\end{abstract}

\vspace{-1mm}
\section{Introduction}
\vspace{-1mm}
Speech coding aims to reduce the bitrate of speech signals while preserving speech quality, which is a fundamental step in applications like speech communication and transmission. 
Conventional speech coding methods have evolved from extensive manual efforts and decades of research \cite{G711,G723,G726,G729}, and have been applied in various practical scenarios. 
However, they still face challenges such as limited speech quality and high bitrates.

Recently, neural speech coding has attracted significant attention. 
Neural speech codecs operate in an end-to-end trainable framework, leveraging neural networks to jointly design and train the encoder, quantizer and decoder. 
SoundStream \cite{zeghidour2021soundstream} represents an early effort in this domain, employing an encoder to directly process the speech waveform, a residual vector quantizer (RVQ) \cite{vasuki2006review} for discretization, and a decoder to reconstruct the speech waveform. 
Subsequent approaches, such as Encodec \cite{defossezhigh} and HiFi-Codec \cite{yang2023hifi}, can be viewed as variants of SoundStream, incorporating more advanced training and quantization strategies. 
To avoid direct waveform modeling and improve generation efficiency, APCodec \cite{ai2024apcodec} employs a parallel encoding and decoding approach for speech amplitude and phase spectra.
More recently, MDCTCodec \cite{jiang2024mdctcodec} has adopted a simpler single-path structure and utilized the modified discrete cosine transform (MDCT) spectrum as the modeling target, achieving higher generation efficiency while maintaining high decoded speech quality.

The aforementioned neural speech codecs rely solely on the speech modality during training, and the model is optimized using frames with a very short window shift \cite{defossezhigh,yang2023hifi,ai2024apcodec,jiang2024mdctcodec} (e.g., 5 ms$\sim$10 ms), which inevitably leads to local optima. 
Intuitively, the decoded speech would be smoother and more coherent if long-term cues were introduced to the codec. 
Ahasan \MakeLowercase{\textit{et al.}} proposed DMCodec \cite{ahasan2024dm}, which incorporates a language model (LM) and self-supervised speech model (SM) into a nerual speech codec. 
It has shown that the semantic and contextual information contained in the LM and SM can enhance the timbre of the decoded speech. 
This indicates that leveraging information from other modalities is beneficial for improving speech coding quality.

Other modalities, such as the visual modality, also contain rich speech-related information, e.g., speaker characteristics. 
Integrating visual information to assist speech generation tasks has gained widespread attention \cite{alfouras2018conversation,xu2022vsegan,zheng2024incorporating,xu2022improving}.
For example,  Xu \MakeLowercase{\textit{et al.}} proposed a multi-layer fusion model with multi-head cross-attention mechanism to fuse audio and lip features for audio-visual speech enhancement in \cite{xu2022improving}. 
Zheng \MakeLowercase{\textit{et al.}} incorporated ultrasound tongue images to improve the performance of lip-based speech enhancement~\cite{zheng2024incorporating}. 
Researchers have also collected audio-visual corpus \cite{ribeiro2021tal} and made efforts towards audio-visual speech recognition \cite{afouras2018deep}. 
However, integrating visual information into neural speech codecs to improve performance has not yet been thoroughly investigated.

\begin{figure}
    \centering
    \includegraphics[width=1\linewidth]{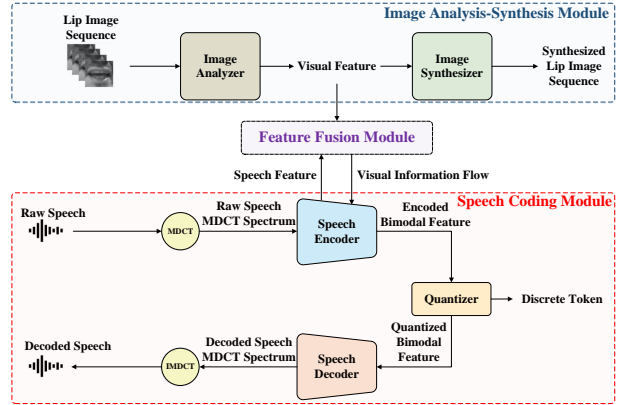}
    \caption{An overview of the proposed VNSC.}
    \label{fig1}
\end{figure}

Therefore, this paper proposes VNSC, a vision-integrated neural speech codec, to explore the potential of visual modality in improving coding quality. 
The image analysis-synthesis module of VNSC extracts visual features from lip images, which are then fully integrated with speech features through the feature fusion module to forms bimodal features. These features are then injected into a standard neural speech codec to assist the coding process. 
For two different application scenarios, we design two distinct feature fusion strategies. 
When visual information is available during inference, visual features are explicitly integrated into the speech coding process through concatenation. In contrast, when visual information is unavailable during inference, visual features are implicitly incorporated into the speech coding process during training through distillation.
Experimental results confirm that with the assistance of visual modality, both the decoded speech quality and the noise robustness of the neural speech codec are significantly improved. 

Finally, we draw conclusions in Section \ref{Conclusion}.

\begin{figure*}
    \centering
    \includegraphics[width=1\linewidth]{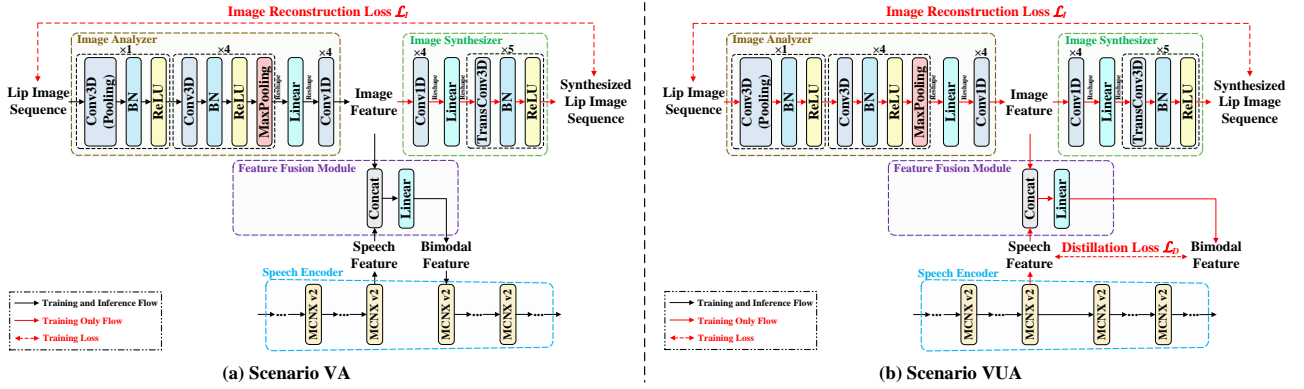}
    \caption{Structural details of the image analysis-synthesis module and the feature fusion module in VNSC. Here, Conv3D, TransConv3D, and Conv1D represent 3D convolution, transposed 3D convolution and 1D convolution operations, respectively. For simplicity, only the MCNX v2 blocks of the speech encoder in the speech coding module is depicted.}
    \label{fig:2}
\end{figure*}

\vspace{-1mm}
\section{Proposed Method}
\vspace{-1mm}
\label{Proposed Method}

\subsection{Overview}

As shown in Figure \ref{fig1}, VNSC consists of a speech coding module, an image analysis-synthesis module and a feature fusion module. 
The image analysis-synthesis module extracts visual features from lip images and injects them into the speech encoding module via the feature fusion module to assist the coding process. The VNSC is designed for two scenarios, including

\begin{itemize}
\item {}{\textbf{Scenario VA(Video Available)}}: Visual information is available during inference. 
The speech coding process can explicitly leverage visual information. 

\item {}{\textbf{Scenario VUA(Video Unavailable)}}: Visual information is unavailable during inference. 
The speech coding process can only implicitly incorporate visual information. 
In this scenario, the speech codec should learn the visual information even in the absence of visual input, offering the advantage of incurring no additional computational during inference.

\end{itemize}

\vspace{-1mm}
\subsection{Speech Coding Module}
\vspace{-1mm}
We use MDCTCodec \cite{jiang2024mdctcodec}, an efficient and lightweight neural speech codec, in the speech coding module. 
The speech encoder takes the MDCT spectrum $\bm{M}\in\mathbb R^{M\times N}$ of the raw speech as input, combines it with visual information from the feature fusion module, and outputs the encoded bimodal features, where $M$ and $N$ represent the dimensions and the number of frames of the MDCT spectrum, respectively. 
These features are then discretized by an RVQ. 
The speech decoder finally decodes the MDCT spectrum from the quantized results and reconstructs the waveform via inverse MDCT (IMDCT).


The speech encoder includes a pre-processing module, eight cascaded modified ConvNeXt v2 (MCNX v2) blocks and a post-processing module. 
At the input end of the speech encoder, a plain 1D convolutional layer and layer normalization are employed for feature pre-processing. At the output end, the post-processing module comprises layer normalization, a linear layer, a downsampling 1D convolutional layer and a plain 1D convolutional layer. 
The MCNX v2 blocks have proven to be well-suited for speech coding \cite{ai2024apcodec,jiang2024mdctcodec}. 
Each MCNX v2 block primarily consists of a 1D depth-wise convolutional layer followed by layer normalization, a linear layer followed by a global response normalization \cite{woo2023convnext} layer and a Gaussian error linear unit activation \cite{hendrycks2016gaussian} function, with residual connections.
The speech decoder is largely mirror-symmetric to the speech encoder, except that the downsampling 1D convolutional layer is replaced by a 1D upsampling transposed convolutional layer.

\subsection{Image Analysis-Synthesis Module}

As shown in Figure \ref{fig:2}, the image analysis-synthesis module consists of an image analyzer $\phi_{IA}$ and an image synthesizer $\phi_{IS}$. 
The image analyzer processes the input lip image sequence $\bm{I} \in {\mathbb{R}^{{H}\times{W} \times N}}$ and extracts the visual feature $\bm{V} \in {\mathbb{R}^{{D_v} \times N}}$, where $H$ and $W$ denote the height and width of each image, respectively, and $D_v$ represent the dimension of the visual feature. 
The image sequence $\bm{I}$ is time-aligned with the MDCT spectrum $\bm{M}$ in the speech coding module, with the temporal length being $N$. 
The image synthesizer then reconstructs the visual feature $\bm{V}$ into the image $\hat{\bm{I}} \in {\mathbb{R}^{{H}\times{W} \times N}}$. 
This process can be represented by the following equations:
\begin{equation}
\bm{V} = \phi_{IA}(\bm{I}),\
\hat {\bm{I}} = \phi_{IS}(\bm{V}).
\end{equation}
The extracted visual feature is sent to the feature fusion module for further processing before being incorporated into the speech coding process. 
For VA scenario, only the image analyzer are used during inference; whereas for VUA scenario, the entire image analysis-synthesis module is absent during inference.

The image analyzer consists of five cascaded analysis blocks and a post-processing block. 
The first analysis block comprises a 3D convolution, batch normalization (BN) and a ReLU activation function. 
This 3D convolution reduces the height and width by applying a stride, effectively performing pooling. 
The last four analysis blocks include an additional pooling layer to reduce the height and width, with the 3D convolution no longer serves this function. 
The post-processing block consists of a linear layer and four 1D convolutional layers, primarily responsible for reshaping the features and reducing the feature dimensions. 
The image synthesizer is largely mirror-symmetric to the image analyzer, with all the 3D convolutions replaced by 3D transposed convolutions, and the pooling layers removed. 
The increase in height and width is achieved by five 3D transposed convolutions.


During the training phase, for both scenarios, we define an image reconstruction loss between $\bm{I}$ and $\hat{\bm{I}}$ as their mean squared error (MSE), i.e., 
\begin{equation}
\mathcal L_{I} = \frac{1}{{HWN}}\mathbb E_{(\bm{I},\hat{\bm{I}})}\left\Vert \bm{I}-\hat{\bm{I}}\right\Vert_F^2,
\end{equation}
where $\Vert\cdot\Vert_F$ denotes the Frobenius norm.


\subsection{Feature Fusion Module}

The feature fusion module is responsible for integrating speech features and visual features, and transmitting the bimodal features into the speech coding process. 
As shown in Figure \ref{fig:2}, feature interaction occurs between the feature fusion module and the speech encoder in the speech coding module. 
Let the output speech feature of the $i$-th MCNX v2 block in the speech encoder be $\bm{X}_i\in\mathbb R^{D_s\times N}$, where $D_s$ represents the dimension of the speech feature. 
The feature fusion module concatenates the speech feature $\bm{X}_i$ and the visual feature $\bm{V}$ along the dimension axes, and applies a linear layer to reduce the dimensionality, ensuring that the bimodal feature dimensions match those of the speech feature, i.e.,
\begin{equation}
    \hat{\bm{X}}_i =  {\rm Concat}\{\bm{X}_i, \bm{V}\}, \
    \tilde {\bm{X}}_i = {\rm Linear}(\hat{\bm{X}}_i),
\end{equation}
where $\hat{\bm{X}}_i\in\mathbb R^{(D_s+D_v)\times N}$ denotes the concatenated feature and $\tilde {\bm{X}}_i\in\mathbb R^{D_s\times N}$ represents the bimodal feature.

As shown in Figure \ref{fig:2}, the bimodal feature integration manner differs significantly between the two scenarios. 
In the VA scenario, since the lip images are available during the inference phase, we adopt an explicit integration approach, directly using the bimodal feature $\tilde {\bm{X}}_i$ as input to the next MCNX v2 block, i.e., the $(i+1)$-th block. 
However, for the VUA scenario, since the lip images are unavailable during the inference phase, we adopt an implicit integration approach, distilling the information from the bimodal feature $\tilde {\bm{X}}_i$ into the speech feature $\bm{X}_i$ during the training phase. 
Specifically, we define the distillation loss $\mathcal L_D$ between $\bm{X}_i$ and $\tilde {\bm{X}}_i$ as
\begin{equation}
\mathcal L_D =\log\left(1 + {e^{ - \frac{{{tr(\bm{X}_i^\top}  {{\tilde{ \bm{X}}}_i)}}}{{\max (||{\bm{X}_i}{||_F},\varepsilon ) \cdot \max (||{{\tilde{\bm{X}}}_i}{||_F},\varepsilon )}}}}\right),
\end{equation}
where $tr$ denotes the matrix trace and $\varepsilon=1e-6$ is a lower bound. 
The distillation loss encourages the speech features produced by the MCNX v2 blocks to learn certain visual information, ensuring that during inference, the speech encoder can still generate features containing visual information even without lip images, which can then be used in the speech coding process.

\subsection{Training Criteria}

The three modules of VNSC are jointly trained using the generative adversarial training strategy provided by MDCTCodec \cite{jiang2024mdctcodec}. 
The speech coding module fully adopts the loss function $\mathcal L_{MDCTCodec}$ from MDCTCodec \cite{jiang2024mdctcodec}, which includes the generative adversarial loss, MDCT spectrum loss, Mel spectrogram loss, and quantization loss. 
For the VA scenario, the loss function of VNSC is defined as a linear combination of $\mathcal L_{MDCTCodec}$ and the image reconstruction loss $\mathcal L_{I}$, i.e., 
\begin{equation}
\begin{aligned}
\mathcal L_{VNSC} = \mathcal L_{MDCTCodec} + \lambda _I \mathcal L_I.
\end{aligned}
\end{equation}
For the VUA scenario, the loss function additionally incorporates the additional distillation loss $\mathcal L_{D}$, i.e.,
\begin{equation}
\begin{aligned}
\mathcal L_{VNSC} = \mathcal L_{MDCTCodec} + \lambda _I \mathcal L_I + \lambda _D \mathcal L_D,
\end{aligned}
\end{equation}
where $\lambda _I$ and $\lambda _D$ are hyperparameters.


\section{Experimental Setup}
\label{Experimental Setup}

\subsection{Datasets}
In the experiment, we utilized the Tongue and Lips (TaL) corpus \cite{ribeiro2021tal}, a multi-speaker dataset containing ultrasound tongue imaging, optical lip videos and speech for each utterance. 
We focused on the TaL80 subset, which included recordings from 81 native English speakers without any voice talent. 
Only the speech and lip video data were used in our experiments. 
The lip video was recorded at a frame rate of 60 fps, and the speech was recorded at a sampling rate of 48 kHz at a depth of 16 bit. 
The training and validation sets contained 11,478 and 810 utterances, respectively, while the remaining 1,140 utterances formed the test set for inference.
The content of these three sets was mutually exclusive. 

\subsection{Experimental Settings}

In VNSC\footnote{Speech samples are available at \url{https://doge114514-bot.github.io/VNSC_demo/}.}, the configuration of the speech coding module is fully borrowed from MDCTCodec with a bitrate of 6 kbps \cite{jiang2024mdctcodec}. 
The frame shift for extracting MDCT spectrum from the speech is set to 40, resulting in a frame rate of 1.2 kHz for the speech feature $\bm{X}_i$. 
The dimension of $\bm{X}_i$ is set to 256 (i.e., $D_s=256$).
For the image analysis-synthesis module, to achieve temporal alignment between speech and visual features, we first upsampled the lip video to an image sequence at 150 Hz using FFmpeg \cite{tomar2006converting}, and then replicated each time point in the sequence eight times to achieve a 1.2 kHz frame rate. 
Each image has a height and width of 64. 
The image sequence is first reshaped into the shape [$H=64$, $W=64$, $C=1$, $N$] before being fed into the image analyzer. 
Here, $C$ represents the number of channels, and $N$ is determined by the length of the speech. 
For the analysis blocks in the image analyzer, the kernel size of all five 3D convolutions is set to 3, and the output channels of these five 3D convolutions are set to 32, 64, 128, 256 and 512, respectively. 
The stride of the first 3D convolution is set to 1 along the time axis, and 2 along the height and width axes, while the stride of the remaining four 3D convolutions is set to 1 for all axes. 
All four pooling operations adopt a 2$\times$2 pooling window for height and width axes. 
Therefore, the feature shape output by the final analysis block is [$H=2$, $W=2$, $C=512$, $N$]. 
Then, the post-processing block first merges the height and width axes of the feature, eliminates this axis through a linear layer, and finally reduces the dimensions through four 1D convolution layers with kernel sizes of 3 and channel numbers of 256, 256, 64, and 64, respectively, resulting in visual feature $\bm{V}$ with a shape of [$C=64$, $N$] (i.e., $F_v=64$).
The configuration of the image synthesizer is essentially the same as that of the image analyzer. 
The number of nodes of the linear layer in the feature fusion module is set to 256.


We trained VNSC on a single Nvidia RTX 4090-D GPU with batch size of 16. 
The hyperparameters were set as $\lambda _I$ = $10^{-5}$ for the VA scenario, and $\lambda _I$ = $0.5\times 10^{-5}$ and $\lambda _D = 1$ for the VUA scenario. 
We employed the AdamW optimizer \cite{loshchilov2018decoupled} for training, with exponential decay rate $\beta _1 = 0.8$ and $\beta _2 = 0.99$. 
The learning rate decayed by a factor of 0.999 after each epoch, starting from an initial rate of 0.0002. 

In the experiments, we compared VNSC with its base model MDCTCodec \cite{jiang2024mdctcodec}, as well as with several other advanced neural speech codecs, including SoundStream\cite{zeghidour2021soundstream}, Encodec\cite{defossezhigh} and HiFi-Codec\cite{yang2023hifi}. 
We used several commonly used objective metrics to evaluate speech quality, including perceptual evaluation of speech quality (PESQ) \cite{rec2007p}, three composite measures (CSIG, CBAK and COVL) \cite{hu2006evaluation}, short-time objective intelligibility (STOI) \cite{taal2010short} and virtual speech quality objective listener (ViSQOL). 
For noise robustness evaluation, segmental signal-to-noise ratio (SSNR) was also used.
For all metrics, higher values indicate better performance.



\section{Experimental Results}
\label{Experimental Results}

\vspace{-1mm}
\subsection{Determination of Feature Integration Locations}
\vspace{-1mm}

\begin{table}[t]
\centering
\caption{Evaluation results of the bimodal feature integration locations and the image reconstruction loss of VNSC on the validation set for VA scenario.}
\label{Table 1}
\setlength{\tabcolsep}{0.4mm}{
\begin{tabular}{c|cccccc}
\hline
      & PESQ         & CSIG         & CBAK         & COVL         & STOI & ViSQOL       \\ \hline
$i=1$   & 3.28          & 4.82          & 3.48          & 4.09          & 0.95  & 3.97          \\
$i=2$   & \textbf{3.33} & \textbf{4.85} & \textbf{3.50} & \textbf{4.13} & 0.95  & \textbf{3.98} \\
$i=3$   & 3.20          & 4.77          & 3.38          & 4.02          & 0.95  & 3.91          \\
$i=4$   & 3.17          & 4.73          & 3.28          & 3.98          & 0.95  & 3.86          \\ \hline
w/o $\mathcal L_I$ $(i=2)$ & 3.22          & 4.78          & 3.41          & 4.04          & 0.95  & 3.94          \\ \hline
\end{tabular}}
\end{table}

We first determined the optimal location for integrating bimodal features in VNSC through experiments, specifically determining the value of $i$ such that the performance is maximized when visual information starts flowing into the speech coding process at the $(i+1)$-th MCNX v2 block. 
The experiments were conducted only for the VA scenario. 

The experimental results on the validation set are shown in Table \ref{Table 1}. 
We can observe that as $i$ increased from 1 to 4, all metrics first increased and then decreased. 
Additionally, we can observe that visual information was integrated too late, the performance degradation is particularly noticeable. 
When $i=2$, VNSC achieved the best performance. 
This may be attributed to the fact that different MCNX v2 blocks in the speech encoder learned various aspects of speech during training. 
The third block, in particular, likely captured features more closely related to the principles of speech production, making the flowing of visual information in this block especially beneficial for learning. 
In the following experiments, VNSC is configured with $i=2$, meaning that visual information flows into the speech coding process starting from the third MCNX v2 block. 


Additionally, we conducted an ablation study on the validation set to confirm the necessity of the image synthesizer and image reconstruction loss $\mathcal L_I$, with the results shown in the last row of Table \ref{Table 1}. 
Clearly, when $\mathcal L_I$ is ablated, all metrics significantly worsen. 
This indicates that the image synthesizer and image reconstruction loss help the image analyzer extract visual information. 
Without them, the speech modality may dominate, and the visual modality would lose its effectiveness.


\begin{table}[t]
\centering
\caption{Evaluation results of VNSC and compared advanced neural speech codecs on the test set.}
\label{Table 2}
\setlength{\tabcolsep}{0.4mm}{
\begin{tabular}{c|cccccc}
\hline
      & PESQ         & CSIG         & CBAK         & COVL         & STOI         & ViSQOL       \\ \hline
VNSC (VA)    & \textbf{3.36} & \textbf{4.87} & 3.51 & \textbf{4.17} & \textbf{0.96} & \textbf{4.01} \\
VNSC (VUA)   & 3.30          & 4.81          & 3.44          & 4.08          & 0.95          & 3.94          \\
MDCTCodec   & 3.26          & 4.79          & 3.29          & 4.07          & 0.95          & 3.92          \\
SoundStream & 2.65             & 4.21             & 3.39             & 3.46             & 0.90             & 3.07             \\
Encodec     & 2.95          & 4.45          & \textbf{3.55}          & 3.73          & 0.92          & 3.21          \\
HiFi-codec  & 3.29          & 4.76          & 3.54          & 4.06          & 0.94          & 3.56          \\ \hline
\end{tabular}}
\end{table}

\subsection{Comparison with Advanced Codecs}

We then compared VNSC with its basic model, MDCTCodec, and other advanced codecs, i.e., SoundStream, Encodec and HiFi-Codec. 
The experimental results on the test set are shown in Table \ref{Table 2}. 
We can see that VNSC for VA scenario significantly outperformed MDCTCodec across almost all metrics, confirming the effectiveness of explicitly incorporating visual information. 
Although VNSC for VUA scenario, like MDCTCodec, did not incorporate visual information during inference, it still showed improvements across all metrics, indicating that it has indeed learned useful visual modality knowledge. 
Compared to other codecs, VNSC effectively bridged the performance gap in certain metrics that MDCTCodec, as a unimodal model, falled short in. 
For example, the CBAK of MDCTCodec showed a large gap compared to HiFi-Codec, but after incorporating visual information to form VNSC, this metric became comparable.
In summary, integrating visual information in VNSC is beneficial for improving speech coding quality, and VNSC considers multiple scenarios, making it practical and versatile.

\vspace{-1mm}
\subsection{Evaluation on Noise Robustness}
\vspace{-1mm}
\begin{table}[t]
\centering
\caption{Noise robustness evaluation results of VNSC and MDCTCodec on a noisy test set.}
\label{Table 3}
\setlength{\tabcolsep}{0.4mm}{
\begin{tabular}{c|ccccccc}
\hline
    & PESQ & CSIG & CBAK & COVL & STOI & ViSQOL & SSNR \\ \hline
VNSC (VA)  & \textbf{2.95} &\textbf{4.34} & \textbf{3.09} & \textbf{3.66} & \textbf{0.94} &\textbf{3.30} & \textbf{2.63}       \\
VNSC (VUA) & 2.90 & 4.26 & 3.05 & 3.60 & 0.94 &3.23 & 2.37       \\
MDCTCodec & 2.85 & 4.25 & 2.97 & 3.56 & 0.93 &3.24 & 1.61      \\ \hline
\end{tabular}}
\end{table}

Finally, we evaluated the noise robustness of VNSC. 
Specifically, we added slight noise to the test set, then encoded the speech using MDCTCodec and VNSC, respectively, and computed the objective metrics by comparing the decoded speech with the original clean speech. 
Experimental results are shown in Table \ref{Table 3}. 
VNSC consistently outperformed MDCTCodec across all metrics in both scenarios. 
The improvement in SSNR was particularly significant, indicating that integrating visual information effectively enhanced the model's noise robustness.


\vspace{-1mm}
\section{Conclusion}
\vspace{-1mm}
\label{Conclusion}

This paper proposed VNSC, a novel bimodal neural speech codec that integrates visual information. 
VNSC is built upon the speech-modal MDCTCodec, with visual information extracted from lip images flowing into the speech coding process. 
For scenarios where lip images may or may not be available during the inference stage, explicit integration and implicit distillation of visual information are employed, respectively. 
Experimental results confirm that the proposed VNSC outperformed the unimodal speech codec in both decoded speech quality and noise robustness. 
Further reducing the latency of VNSC and developing a streamable bimodal codec will be our future work.


\bibliographystyle{IEEEtran}
\bibliography{mybib}

\end{document}